# Carbon Nanomaterials to Combat Virus: A perspective in view of COVID-19

*Joydip Sengupta[1] and Chaudhery Mustansar Hussain[2*]*

[1]Department of Electronic Science Jogesh Chandra Chaudhuri College (Affiliated to University of Calcutta), Kolkata - 700 033, W.B, India;

[2]Department of Chemistry and Environmental Science, New Jersey Institute of Technology, Newark, New Jersey, USA

*Corresponding author Email: chaudhery.m.hussain@njit.edu

**Abstract**

The rapid outbreaks of lethal viruses necessitate the development of novel antiviral substance. Besides the conventional antiviral substances, biocompatible nanomaterials also have significant potential in combating the virus at various stages of infection. Carbon nanomaterials have an impressive record against viruses and can deal with many crucial healthcare issues. In accordance with the published literature, biocompatible carbon nanomaterials have a promising prospect as an antiviral substance. Subsequently, the antiviral properties of different carbon nanomaterials namely fullerene, carbon nanotube, carbon dot and graphene oxide have been reviewed.

**Keywords**: Fullerene; Carbon nanotube; Carbon dot; Graphene oxide; Antiviral; COVID-19.

The world is facing an unprecedented medical emergency since December 2019 owing to a new strain of coronavirus which is named as severe acute respiratory-related coronavirus 2 (SARS-CoV-2)[1]. Various detection strategies are being employed to refrain off the transmission chain of the disease which is mainly based on immunological assays or using amplification techniques. The methods involving immunological assays suffer from the drawback of the requirement of complex production of antibodies and recombinant proteins while amplification-based techniques require expensive instruments and skilled personnel[2]. To overcome the inherent problems of the standard detection techniques many new nanotechnology-based approaches have evolved[3,4] incorporating carbon nanomaterials[5]. However, mere detection is not enough to break the chain, the development of a new generation of antiviral tools is urgently needed to combat this pandemic.

Antiviral substances can be categorised as virostatic and virucidal[6]. The working of virostatic substances is based on a binding mechanism and they act in the early stages of infection by reducing viral replication rate, while virucidal substances can deactivate the virus permanently. These antiviral substances are of the order of nanometer and they interact well with nanosized viruses via biochemical interaction. In this regard, carbon nanomaterials show great potential to be used as an antiviral substance because of their novel physio-chemical characteristic[7].

The first carbon nanomaterial fullerene was discovered in 1985 by Prof Smalley's group and it is also the first candidate to be tested as the antiviral substance in 1993 by Sijbesma et al.[8]. It was found that fullerenes fit well into the active sites of the human immunodeficiency virus (HIV) and can block the encoded enzymes. Later on, several researchers have synthesised different tailor-made fullerene derivatives to achieve higher inhibition of HIV-1 maturation[9], HIV-RT[10] or HIV-1 protease[11]. In some cases, the solubility of antiviral substance is required and interestingly it is possible to synthesise water-soluble fullerene while keeping their antiviral activity intact[12]. Kornev et al.[13] developed water-soluble polycarboxylic derivatives of fullerene which showed pronounced antiviral activity against HIV and influenza A virus. Carboxylic derivatives of fullerene also have been effective to inhibit Herpes Simplex Virus and Cytomegalovirus[14]. Kataoka et al.[15] synthesised proline-type fullerene to inhibit Hepatitis C virus protease successfully. Falynskova et al.[16] produced a tailor-made fullerene compound fullerene-(tris-aminocaproic acid) hydrate to hinder the Respiratory Syncytial Virus multiplication in HEp-2 cell, however application time is crucial for effective inhibition of the virus. Fullerene derivatives are also able to inhibit the Zika Virus and Dengue viruses[17]. Non-derivatized fullerene is also employed for efficient inhibition of viruses like simian immunodeficiency virus and the Moloney murine leukemia virus (M-MuLV)[18]. Moreover, fullerene can produce singlet oxygen ($^1O_2$) upon illumination by visible light[19] (Fig 1). This property of fullerene has been utilized by several researchers for the inactivation of viruses[20,21] even in water[22] and can be used for the sanitization of indoor air[23]. Thus, fullerene can be effectively employed to combat RNA type viruses.

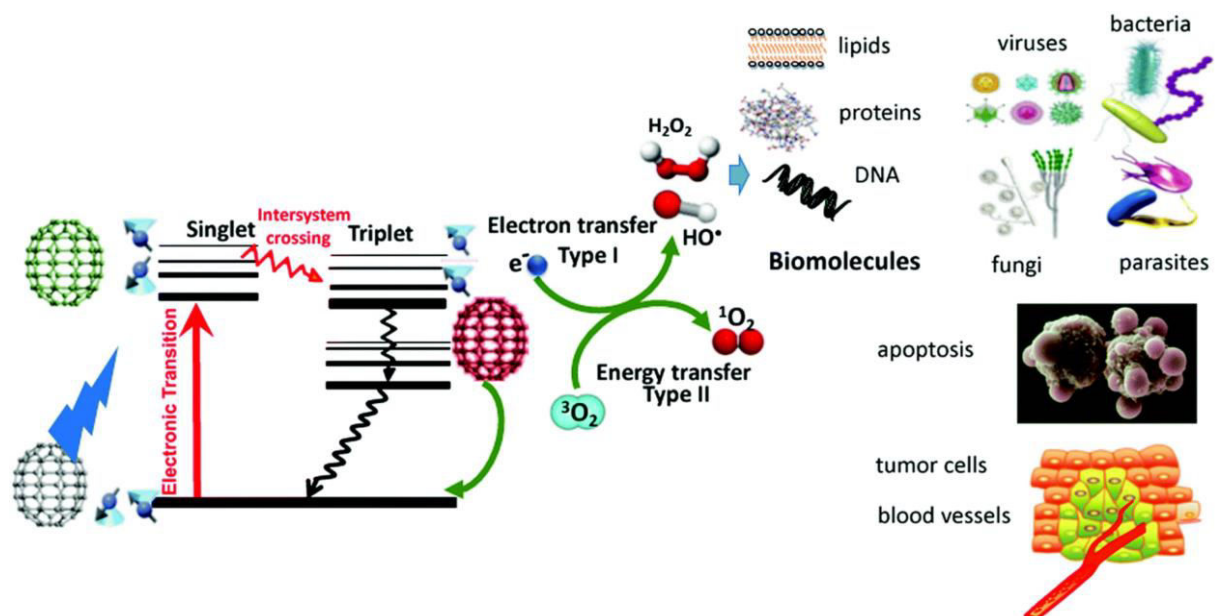

Fig 1. A ground-state fullerene absorbs a photon, transitions to the short-lived (nsec) excited singlet state that can undergo intersystem crossing to the long-lived (µsec) excited triplet state. The triplet fullerene can relax to ground state by emitting phosphorescence, but can also undergo energy transfer with ground state triplet oxygen ($^3O_2$) to form reactive singlet oxygen ($^1O_2$, Type 2) or else can undergo an electron transfer reaction to form HO•, superoxide and $H_2O_2$ (Type 1). These ROS ($^1O_2$ and HO•) can damage lipids, proteins and nucleic acids leading to the destruction of all types of microbial cells, and efficiently kill cancer cells and destroy tumors. (Reproduced with permission from Royal Society of Chemistry[24])

Another carbon allotrope, namely carbon nanotube (CNT) was first reported in 1952[25], but after the description provided by Ijima in 1991,[26] it gains worldwide attention. Krishnaraj et al.[27] theoretically studied the molecular interaction between CNT and HIV. Their results revealed that CNT has a high binding affinity towards the proteins of HIV, demonstrating the antiviral activity. Single wall carbon nanotubes (SWCNTs), both in pristine form and decorated with metal (Pt, Pd, Ni, Cu, Rh, or Ru) nanoparticles were examined to study the adsorption of hydrogen peroxide (as a representative of the ROS family) by Aasi et al.[28] employing density functional theory. The results exhibited that hydrogen peroxide is weakly physisorbed on pristine SWCNT while strong physisorption was found on metal decorated SWCNTs. Moreover, metal decorated SWCNTs showed long shelf lives which make it a potential candidate for designing the antiviral surfaces and personal protection equipment (PPE). Banerjee et al.[29] prepared porphyrin-conjugated multi-walled carbon nanotubes (MWCNTs) by attaching protoporphyrin IX to acid-functionalized MWCNTs. Under the illumination of visible light, the prepared material considerably suppressed the ability of Influenza A virus to infect mammalian cells. Differently functionalized MWCNTs were studied by Iannazzo et al.[30] for their in-vitro anti-HIV activity. The results revealed that hydrophilic functionality and water dispersibility of functionalized MWCNTs governed the antiviral activity against HIV. Henceforth, it can be seen that CNT can be potentially used for deactivation of RNA type viruses.

Carbon dots (CDs) were accidentally discovered by Xu et al.[31] in 2004 and later in 2006 Sun et al. produced stable photoluminescent carbon nanoparticles having different sizes and named as "Carbon Quantum Dots" (CQDs)[32]. On the other hand graphene quantum dot (GQD) was discovered in 2008 by Ponomarenko et al.[33]. There is a structural difference between GQD and CD. CDs are generally quasi-spherical in shape and are smaller than 10 nm in size, while GQD has graphene lattice having a size of less than 100 nm[34]. Barras et al.[35] reported that functionalized CDs with boronic acid can restrict the entry of herpes simplex virus type 1 (HSV-1) at its early stage. Functionalized CDs with boronic acid also have the potential to inhibit HIV-1 entry[36]. Du et al.[37] examined inhibition characteristics of CDs on RNA and DNA model type viruses. Pseudorabies virus (PRV) was considered as a DNA model type virus and porcine reproductive and respiratory syndrome virus (PRRSV) was used as an RNA model type virus for this study. They found that CDs induces interferon-α and inhibit virus replication for both RNA and DNA type viruses (Fig 2). Glycyrrhizic acid was used to synthesis CD by Tong et al.[38] and the prepared CDs depicted a broad spectrum of antiviral activity against various viruses like PRRSV, pseudorabies virus (PRV) and porcine epidemic diarrhea virus (PEDV). Lin et al.[39] derived CDs from curcumin (Cur-CQDs) through one-step dry heating and found that the obtained CDs possessed strong antiviral activity against enterovirus 71 (EV71) via scavenging free radicals and strong antioxidant activity. To examine the antiviral property of curcumin derived CDs for inhibition of coronavirus model type virus Ting et al.[40] used PEDV. It was revealed that the curcumin derived CDs inhibited the viral entry by changing the structure of virus surface proteins and reduced the viral replication via stimulation of the production of interferon-stimulating genes (ISGs). The effect of functionalization using boronic acid on seven types of CDs to inhibit human coronavirus was studied by Łoczechin et al.[41]. The derived carbon nanostructures depicted concentration-dependent virus inactivation at the viral replication step. Huang et al.[42] prepared benzoxazine monomer derived CDs and showed that the derived nanostructure possessed broad-spectrum antiviral ability against life-threatening flaviviruses (Japanese encephalitis, Zika, and dengue viruses) and non-enveloped viruses (porcine parvovirus and adenovirus-associated virus). Norovirus can also be inhibited by surface passivated CDs as reported by Dong et al.[43]. Ju et al.[44] reported that CDs are even effective for the inhibition of viral microRNA by regulating the proliferation and survival of virus-induced cancer cells.

GQD has also has been used for inhibition of virus. Lannazzo et al.[45] synthesized water-soluble GQD by prolonged oxidation of multiwall carbon nanotube and used it for the inhibition of HIV RT.

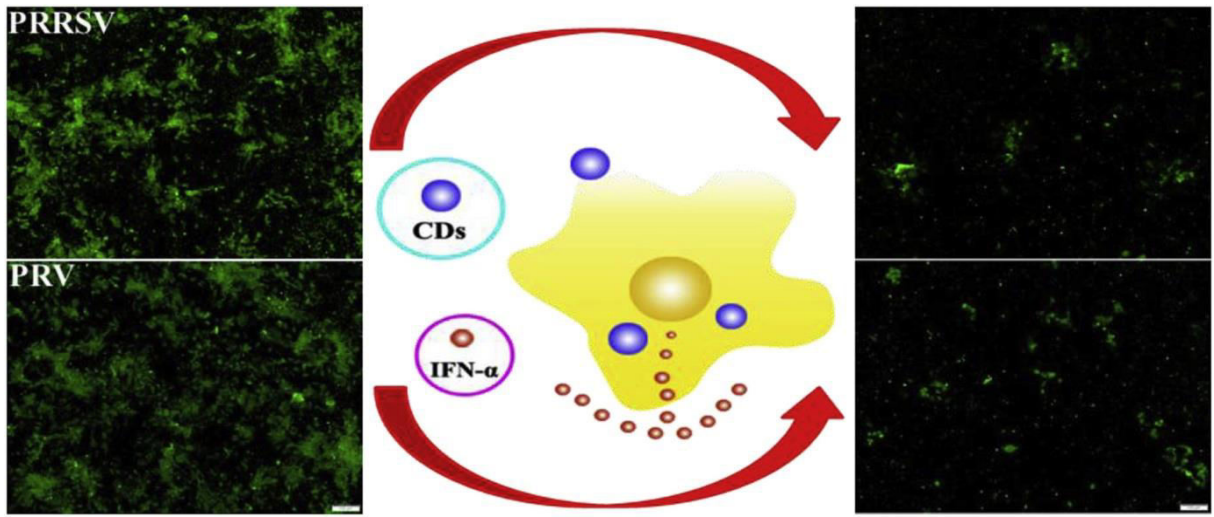

Fig 2. Effect on PRRSV and PRV before (leftmost column) and after (rightmost column) the addition of carbon dots. (Reproduced with permission from Elsevier[46])

After the first demonstration of easy isolation of graphene by Geim and Novoselov, from bulk graphite in 2004[47], graphene and its derivatives have been widely used in different sectors of the medical industry (Fig 3). Sametband et al.[48] used graphene oxide (GO) derivatives to inhibit HSV-1 via viral attachment blocking. The GO blocked HSV-1 infections at relatively low concentrations and the charge density was the major factor affecting the inhibition of virus. Two typical enteric viruses, EV71 and Influenza A virus subtype (H9N2), were used as target viruses by Song et al.[49] for inhibition by GO and found that GO could completely remove the target viruses by destruction and disinfection properties. Yang et al.[50] functionalised GO with β-cyclodextrin and used it as an antiviral substance to inhibit RNA type respiratory syncytial virus. His result showed that the functionalised GO can inactivate the virus directly. The photocatalytic activity of GO[51] has also been explored by the researchers for the inhibition of viruses. Hu et al.[52] prepared GO functioned aptamer and used it for successful photocatalysis of viruses via energy transfer under visible light illumination. Graphene-tungsten oxide composite thin films were synthesized by Akhavan et al.[53] which exhibited an outstanding photocatalytic performance in photoinactivation of bacteriophage MS2 viruses in the presence of visible light.

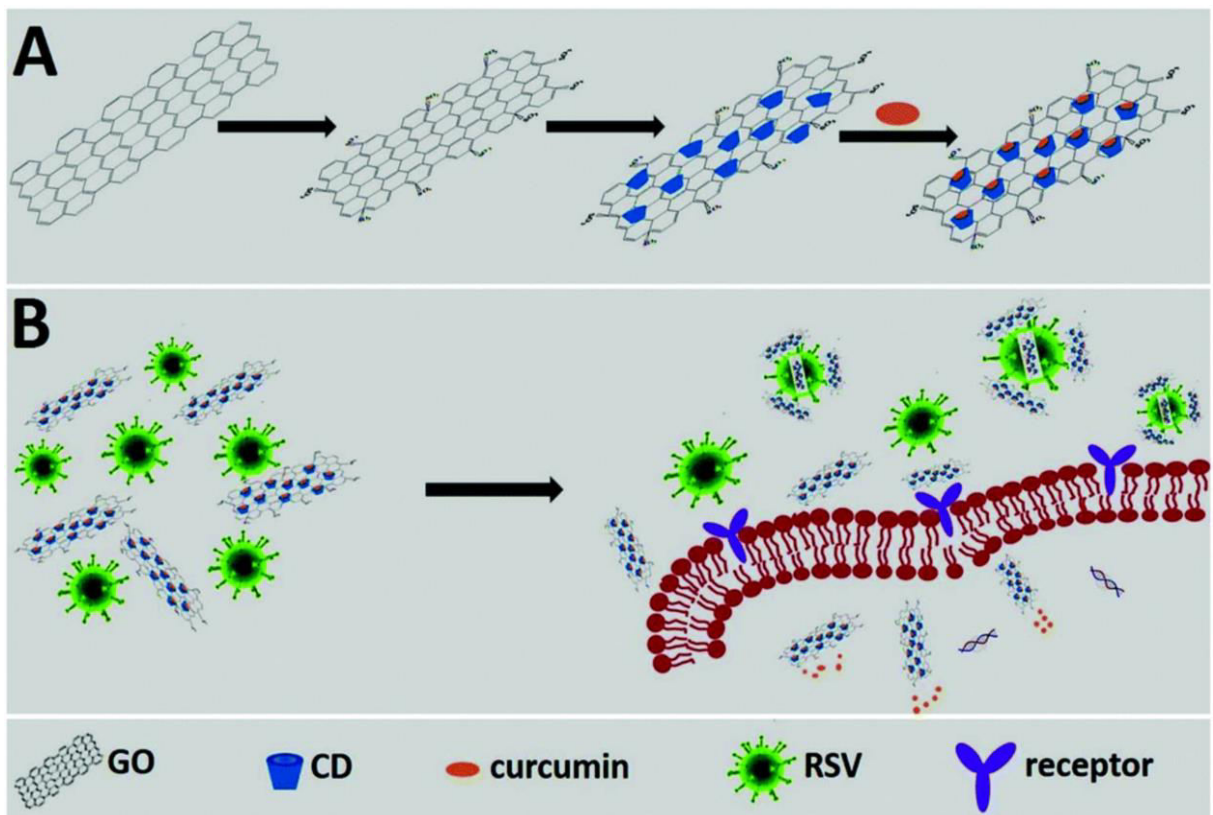

Fig 3. Schematic representation of the working principle. (A) The synthesis of functional nanomaterial composite; (B) the proposed inhibition mode of functional nanomaterials composite against RSV infection. (Reproduced with permission from Royal Society of Chemistry [54])

In summary, SARS-CoV-2 belongs to the RNA virus family and has created a medical emergency worldwide because of its deadly nature and rapid transmission rate. To combat the present situation various carbon nanomaterials, like fullerene, CNT, CD, GQD and GO can be employed as antiviral substances because of their capability of inhibiting RNA type virus[55], biocompatibility and low toxicity. Thus, it can be believed that these unique carbon nanomaterials will pave the way to combat the fatal SARS-CoV-2 in near future.